\documentclass[final,1p,times]{elsarticle} 
\usepackage{graphicx} 
\usepackage{amssymb} 
\usepackage{amsthm} 
\usepackage{lineno} 

\journal{Nuclear Physics A} 
\begin{document} 

\begin{frontmatter} 


\title{Critical point in finite density lattice QCD by canonical approach}

\author{Shinji Ejiri}

\address{Physics Department, Brookhaven National Laboratory, 
Upton, NY 11973, USA}

\begin{abstract} 

We propose a method to find the QCD critical point at finite density calculating the canonical partition function ${\cal Z}_{\rm C} (T,N)$ by Monte-Carlo simulations of lattice QCD, and analyze data obtained by a simulation with two-flavor p4-improved staggered quarks with pion mass  
$m_{\pi} \approx 770 {\rm MeV}$.
It is found that the shape of an effective potential changes gradually as the temperature decreases and a first order phase transition appears in the low temperature and high density region.
This result strongly suggests the existence of the critical point in the $(T, \mu_q)$ phase diagram.

\end{abstract} 

\end{frontmatter} 



\section{First order phase transition and canonical partition function}
\label{sec:cononi}

The critical point terminating a first order phase transition line in the phase diagram of QCD at high temperature and density is one of the most characteristic features that may be discovered in heavy-ion collision experiments. 
To understand the phase structure, first principle calculations of QCD by numerical simulations are very important.
One of the interesting approaches to find a first order phase transition is to construct the canonical partition function ${\cal Z}_{\rm C} (T,N)$ by fixing the total quark number $(N)$ or quark number density $(\rho)$. 
From the canonical partition function, one can estimate the quark number giving the largest contribution to the grand partition function ${\cal Z}_{\rm GC} (T,\mu_q)$. 
Because two different states coexist at a first order transition point, 
two different quark numbers give equally large contributions simultaneously if the transition is of first order. 

The canonical partition function is defined by a fugacity expansion of 
${\cal Z}_{\rm GC} (T,\mu_q)$,
\begin{eqnarray}
{\cal Z}_{\rm GC}(T,\mu_q) 
= \int {\cal D}U \left( \det M(\mu_q/T) \right)^{N_{\rm f}} e^{-S_g}
= \sum_{N} \ {\cal Z}_{\rm C}(T,N) e^{N \mu_q/T}, 
\label{eq:cpartition} 
\end{eqnarray}
where $\det M$, $S_g$ and $N_{\rm f}$ are the quark determinant, the gauge action and the number of flavor, respectively. 
The term ${\cal Z}_{\rm C}(T,N) e^{N \mu_q/T}$ can be regarded as the probability distribution of the quark number $N$.
Moreover, it is worth introducing an effective potential $V_{\rm eff}$ as 
a function of $N$, 
\begin{eqnarray}
V_{\rm eff}(N,T,\mu_q) 
\equiv - \ln {\cal Z}_{\rm C}(T,N) -N \frac{\mu_q}{T} 
= \frac{f(T,N)}{T} -N \frac{\mu_q}{T} , \hspace{3mm}
{\cal Z}_{\rm GC}(T,\mu_q) 
= \sum_{N} \ e^{-V_{\rm eff}}, 
\label{eq:effp} 
\end{eqnarray}
where $f$ is the Helmholtz free energy. 
In a first order phase transition region, 
this effective potential has minima at more than one value of $N$. 
At the minima, the derivative of $V_{\rm eff}$ vanishes: 
\begin{eqnarray}
\frac{\partial V_{\rm eff}}{\partial N} (N, T, \mu_q ) 
= -\frac{\partial (\ln {\cal Z}_{\rm C})}{\partial N} (T,N) 
- \frac{\mu_q}{T} =0 .
\label{eq:dpotdn} 
\end{eqnarray}
Hence, in the first order transition region, we expect 
$\partial (\ln {\cal Z}_{\rm C})/ \partial N (T,N) \equiv - \mu_q^*/T$ 
takes the same value at different $N$. 
Here, $\mu_q^* (T,N)$ is the chemical potential which gives the effective potential at a minimum point $(T, N)$ and becomes $\mu_q$ in the thermodynamic limit.

\begin{figure}[t]
\begin{center}
\includegraphics[width=3.7in]{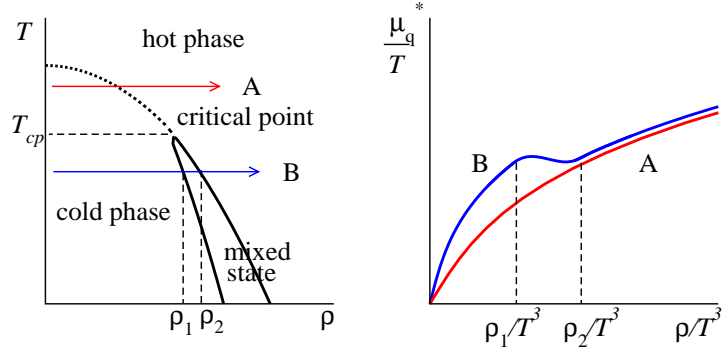}
\vskip -0.2cm
\caption{Phase structure in the $(T, \rho)$ plane and the behavior of 
$\mu_q^*/T$ as a function of $\rho$.
}
\label{fig:cano}
\end{center}
\vskip -0.3cm
\end{figure}

The phase structure in the $(T, \rho)$ plane is sketched in 
the left panel of Fig.~\ref{fig:cano}. 
The thick line is the phase transition line. 
We expect that the transition is crossover at low density and becomes 
of first order at high density. 
In the first order transition region, the two coexisting states are mixed.
The region between the two thick lines is the mixed state. 
The expected behavior of $\mu_q^*$ along the lines A and B are shown 
in the right figure. 
When the temperature is higher than the temperature at the critical 
point $T_{pc}$ (line A), 
$\mu_q^*$ increases monotonically as the density increases. However, 
for the case below $T_{cp}$ (line B), this line crosses the mixed state. 
Corresponding to the double-well potential in a finite volume, 
$\mu_q^*$ is expected to be an S-shaped function.
In the infinite volume limit,
$\mu_q^*$ does not increase in the region between $\rho_1$ and $\rho_2$, 
since the surface energy between the two states can be ignored. 

The Glasgow method \cite{Gibbs86} has been a well-known method to compute the canonical partition function.
A few years ago, the above mentioned behavior at a first order phase transition was observed 
in 4 flavor QCD with staggered fermions  calculating the quark determinant 
by the Glasgow algorithm on a small lattice \cite{Krat05}.
Also, simulations with a fixed quark number, i.e. in the canonical ensemble, have been tried \cite{Li08}.
However, the calculations by the Glasgow algorithm and the simulations for a canonical ensemble require large computational cost and are difficult except on a small lattice with present day computer resources.
In this report, we propose a method based on a saddle point approximation \cite{eji08}. By this approximation, the computational cost is drastically reduced and the first order like behavior was observed for 2 flavor QCD.

\section{Inverse Laplace transformation with a saddle point approximation}

From Eq.~(\ref{eq:cpartition}),
the canonical partition function can be obtained by an inverse Laplace 
transformation,
\begin{eqnarray}
{\cal Z}_{\rm C}(T,N) = \frac{3}{2 \pi} \int_{-\pi/3}^{\pi/3} 
e^{-N (\mu_0/T+i\mu_i/T)} {\cal Z}_{\rm GC}(T, \mu_0+i\mu_i) \ 
d \left( \frac{\mu_i}{T} \right) ,
\label{eq:canonicalP} 
\end{eqnarray}
where $\mu_0$ is an appropriate real constant and $\mu_i$ is 
a real variable. We have used the fact that 
${\cal Z}_{\rm GC}(T, \mu_q +2\pi iT/3) = {\cal Z}_{\rm GC}(T, \mu_q)$
for any complex $\mu_q$.
The grand partition function can be evaluated by 
the following expectation value at $\mu_q=0$.
\begin{eqnarray}
\frac{{\cal Z}_{\rm GC}(T, \mu_q)}{{\cal Z}_{\rm GC}(T,0)}
\hspace{-3mm} &=& \hspace{-3mm} 
\frac{1}{{\cal Z}_{\rm GC}} \int {\cal D}U 
\left( \frac{\det M(\mu_q/T)}{\det M(0)} \right)^{N_{\rm f}}
\det M(0) ^{N_{\rm f}} e^{-S_g} 
= \left\langle 
\left( \frac{\det M(\mu_q/T)}{\det M(0)} \right)^{N_{\rm f}}
\right\rangle_{(T, \mu_q=0)}. 
\label{eq:normZGC} 
\end{eqnarray}

We apply a saddle point approximation to evaluate Eq.~(\ref{eq:canonicalP}), 
which reduces the computational cost. 
If one selects the $\mu_0$ at a saddle point in Eq.~(\ref{eq:canonicalP}), the necessary information is limited that around the saddle point when the volume is sufficiently large. 
Moreover, if we restrict ourselves to study the low density region, the value of $\det M (\mu_q/T)$ near the saddle point can be estimated by a Taylor expansion around $\mu_q=0$. 
The calculation by the Taylor expansion is much cheaper than the exact calculation and the study using a large lattice is possible. 
Also, the truncation error can be systematically controlled by increasing the number of the expansion coefficients.

We assume the existence of a saddle point $z_0$ in the complex $\mu_q/T=z$ 
plane for each configuration, which satisfies
$D'(z_0) - \bar{\rho} =0$. 
Here, the quark number density in a lattice unit and a physical unit are
$\bar{\rho}=N/N_s^3$ and $\rho/T^3=\bar{\rho} N_t^3$, respectively, 
$(\det M(z)/\det M(0))^{N_{\rm f}} = \exp[N_s^3 D(z)] $ and 
$D'(z)=dD(z)/dz$.
We then perform a Taylor expansion around the saddle point and obtain 
the canonical partition function, 
\begin{eqnarray}
{\cal Z}_{\rm C}(T, \bar{\rho} V) 
&=& \frac{3}{2 \pi} {\cal Z}_{\rm GC}(T,0) \left\langle \int_{-\pi/3}^{\pi/3} 
\exp \left[ V \left(D (z_0) - \bar{\rho} z_0 
- \frac{1}{2} D'' (z_0) x^2 + \cdots \right) \right] 
dx \right\rangle_{(T, \mu_q=0)} \nonumber \\
& \approx & \frac{3}{\sqrt{2 \pi}} {\cal Z}_{\rm GC} (T,0)
\left\langle \exp \left[ V \left( D(z_0) - \bar{\rho} z_0 \right) \right] 
e^{-i \alpha/2} \sqrt{ \frac{1}{V |D''(z_0)|}}
\right\rangle_{(T, \mu_q=0)} ,
\label{eq:zcspa}
\end{eqnarray}
where $ D''(z) = d^2 D(z) /dz^2,$ $V \equiv N_s^3$ and 
$D''(z)=|D''(z)| e^{i \alpha}$.
Higher order terms in the expansion of $D(z)$ are negligible when 
the volume $V$ is sufficiently large. 

Within the framework of the saddle point approximation, 
the derivative of $\ln {\cal Z}_C$ with respect to $N$ or $\rho$ 
can be evaluated by 
\begin{eqnarray}
\frac{\mu_q^*}{T} 
= - \frac{1}{V} \frac{\partial \ln {\cal Z}_C (T, \bar{\rho} V)}
{\partial \bar{\rho}}
\approx \frac{
\left\langle z_0 \ \exp \left[ V \left( D(z_0)
- \bar{\rho} z_0 \right) \right] 
e^{-i \alpha /2} \sqrt{ \frac{1}{V |D''(z_0)|}}
\right\rangle_{(T, \mu_q=0)}}{
\left\langle \exp \left[ V \left( D(z_0) 
- \bar{\rho} z_0 \right) \right] 
e^{-i \alpha /2} \sqrt{ \frac{1}{V |D''(z_0)|}}
\right\rangle_{(T, \mu_q=0)}}. 
\label{eq:chemap}
\end{eqnarray}
This formula is similar to that of the reweighting method 
for finite $\mu_q$. 
The operator in the denominator corresponds to a reweighting factor, 
and $\mu_q^* /T$ is an expectation value of the saddle 
point calculated with this modification factor.

\begin{figure}[t]
\begin{center}
\includegraphics[width=2.7in]{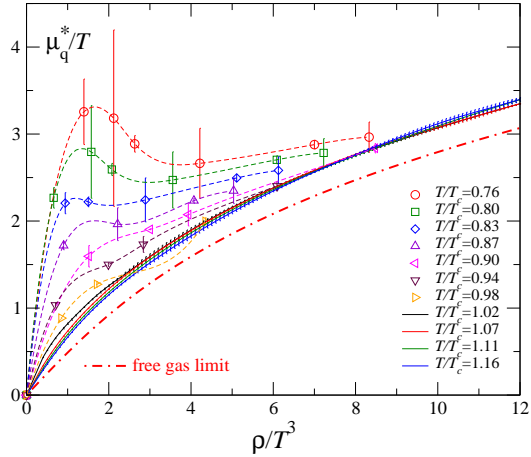}
\vskip -0.2cm
\caption{Chemical potential vs. quark number density for $N_f=2$ 
with a saddle point approximation.
}
\label{fig:chem}
\end{center}
\vskip -0.3cm
\end{figure}

\section{Numerical results and conclusions}

We compute the derivative of $\ln {\cal Z}_C$ using the data obtained in \cite{BS05} with the 2 flavor p4-improved staggered quark action, $m_{\pi} \approx 770 {\rm MeV}$. 
Because the operators in Eq.~(\ref{eq:chemap}) are complex, this calculation suffers from the sign problem. To eliminate the sign problem, the approximation proposed in \cite{eji07} is used: If one assumes that the distribution of the complex phase is well-approximated by a Gaussian function, the complex phase factor $e^{i \theta}$ can be replaced by $\exp[- \langle \theta^2 \rangle /2]$. 
We estimate the quark determinant by the Taylor expansion up to $O(\mu_q^6)$. 
Because the calculation of Eq.~(\ref{eq:chemap}) is similar to the calculation by the reweighting method, the configurations which give important contribution are changed by the modification (reweighting) factor. 
To avoid this problem, we use the multi-$\beta$ reweighting method. 
By this method, the important configurations are automatically selected among all configurations generated at many simulation points of $(T, \mu_q=0)$.
The details are given in \cite{eji08}.

The result of $\mu_q^*/T$ is shown in Fig.~\ref{fig:chem} as a function of $\rho/T^3$ for each temperature $T/T_c$. 
$T_c$ is the pseudo-critical temperature at $\mu_q=0$.
The dot-dashed line is the value of the free quark-gluon gas in 
the continuum theory, 
$\rho/T^3 = N_{\rm f} [ (\mu_q/T) + (1/\pi^2) (\mu_q/T)^3]$.
From this figure, we find that a qualitative feature of $\mu_q^*/T$ 
changes around $T/T_c \sim 0.8$, i.e. $\mu_q^*/T$ increases monotonically 
as $\rho$ increases above 0.8, whereas it shows an S-shape below 0.8. 
This S-shape is a signature of a first order phase transition. 
With some approximations, the critical value of $\mu_q^*/T$ is is estimated to be about $2.4$, which is roughly consistent with the critical point estimated in \cite{eji08} by calculating the effective potential of the plaquette using the same configurations, $(T/T_c, \mu_q/T) \approx (0.76, 2.5)$. 
The difference between these two results may include a systematic error. 
Our result strongly suggests the existence of the critical point terminating the first order phase transition line in the $(T, \mu_q)$ phase diagram of QCD.
Further studies are necessary to predict the critical point quantitatively. 

\paragraph{Acknowledgments}
This work has been authored under Contract No.~DE-AC02-98CH10886
with the U.S. Department of Energy.


\end{document}